%% file: main.tex
\documentclass[runningheads]{llncs}
\usepackage{cite}
\usepackage{amsfonts}
\usepackage{amsmath}
\usepackage{amssymb}
\usepackage{algorithm}
\usepackage{algorithmic}
\usepackage{graphicx}
\usepackage{textcomp}
\usepackage{xcolor}
\usepackage{hyperref}
\hypersetup{
 colorlinks=true,
 citecolor=black,
 linkcolor=black,
 urlcolor=blue}

\usepackage{listings}
\usepackage{float}
\usepackage[mode=buildnew]{standalone}
\usepackage{tikz,pgfplots}
\usetikzlibrary{automata,arrows}
\usetikzlibrary{calc,trees,positioning,arrows,chains,shapes}
\usetikzlibrary{shapes.geometric}
\usetikzlibrary{shapes.multipart}
\definecolor{darkgreen}{rgb}{0,0.6,0}\usepackage{tikz,pgfplots}
\usetikzlibrary{automata,arrows}
\usetikzlibrary{calc,trees,positioning,arrows,chains,shapes}
\usetikzlibrary{shapes.geometric}
\usetikzlibrary{shapes.multipart}
\definecolor{darkgreen}{rgb}{0,0.6,0}

\newfloat{lstfloat}{htbp}{lop}
\floatname{lstfloat}{Listing}

\lstdefinestyle{Python}{
    language        = Python,
    basicstyle      = \scriptsize\ttfamily,
    keywordstyle    = \color{blue},
    keywordstyle    = [2] \color{teal}, 
    stringstyle     = \textcolor{green},
    commentstyle    = \color{red}\ttfamily,
    emphstyle       = \color{blue},
}

\DeclareMathOperator{\Tr}{Tr}

\def\BibTeX{{\rm B\kern-.05em{\sc i\kern-.025em b}\kern-.08em
    T\kern-.1667em\lower.7ex\hbox{E}\kern-.125emX}}
\begin{document}
\lstset{
    showspaces  = false,
    showstringspaces    = false,
    emph={jnp., sum,cos}
}

\title{Reducing Memory Requirements of Quantum Optimal Control\thanks{ This material is based upon work supported by the U.S. Department of Energy, Office of Science, Office of Advanced Scientific Computing Research, under the Accelerated Research in Quantum Computing and Applied Mathematics programs, under contract DE-AC02-06CH11357, and by the National Science Foundation Mathematical Sciences Graduate Internship. We gratefully acknowledge the computing resources provided on Bebop and Swing, a high-performance computing cluster operated by the Laboratory Computing Resource Center at Argonne National Laboratory.}}

\author{
Sri Hari Krishna Narayanan\inst{1} \and
Thomas Propson\inst{2}  \and
Marcelo Bongarti\inst{3} \and
Jan H\"{u}ckelheim\inst{1} \and
Paul Hovland\inst{1}
}
\institute{
Argonne National Laboratory, Lemont, IL 60439, USA \\
\email{\{snarayan, jhueckelheim,  hovland\}@anl.gov}
\and
University of Chicago, Chicago, IL, USA \\
\email{tcpropson@uchicago.edu}
\and
Weierstrass Institute for Applied Analysis and Stochastics, Berlin, Germany \\
\email{bongarti@wias-berlin.de}
}

\authorrunning{S.H.K. Narayanan et al.}
%
\maketitle              
\begin{abstract}
Quantum optimal control problems are typically solved by gradient-based algorithms such as GRAPE, which suffer from exponential growth in storage with increasing number of qubits and linear growth in memory requirements with increasing number of time steps.
These memory requirements are a barrier for simulating large models or long time spans. We have created a nonstandard automatic differentiation technique that can compute gradients needed by GRAPE by exploiting the fact that the inverse of a unitary matrix is its conjugate transpose. Our approach significantly reduces the memory requirements for GRAPE, at the cost of a reasonable amount of recomputation. We present benchmark results based on an implementation in JAX.

\end{abstract}

\section{Introduction}

Quantum computing is computing using quantum-mechanical phenomena,
such as superposition and entanglement. It holds
the promise of being able to efficiently solve problems that
classical computers practically cannot.
In quantum computing,
quantum algorithms are often expressed by using a quantum circuit model,
in which a computation is a sequence of quantum gates.
Quantum gates are the building blocks of quantum circuits and operate on a small number of qubits,
similar to how classical logic gates operate on a small number of bits in conventional digital circuits.

Practitioners of quantum computing must map the logical quantum gates
onto the physical quantum devices that implement quantum gates
through a process called {\em quantum control}. The goal of quantum control
is to actively manipulate dynamical processes
at the atomic or molecular scale, typically by means of
external electromagnetic fields. The objective of {\em quantum optimal
control} (QOC) is to devise and implement shapes of pulses of external fields or
sequences of such pulses that reach a given task in a quantum
system in the best way possible.

We follow the QOC model presented in ~\cite{PhysRevA.95.042318}. Given an intrinsic Hamiltonian $H_0$, an initial state $|\psi_0\rangle$, and a set of control operators
$H_1, H_2, \ldots H_m$, one seeks to determine, for a sequence of time steps
$t_0, t_1, \ldots, t_N$, a set of control fields $u_{k,j}$ such that

\begin{eqnarray}
\mathbb{H}_j & = & H_0 + \sum_{k=1}^{m}u_{k,j}H_k \label{evovlveschrodingersicrete1}\\
U_j & = & e^{-i\mathbb{H}_j(t_j-t_{j-1})}\label{evovlveschrodingersicrete2}\\
K_j & = & U_{j}U_{j-1}U_{j-2} \ldots U_{1}U_{0}\\
|\psi_j\rangle & = & K_j|\psi_0\rangle. \label{evovlveschrodingersicrete4}
\end{eqnarray}

An important observation is that the dimensions of $K_j$ and $U_j$ are  $2^q \times 2^q$, where $q$ is the
number of qubits in the system.
One possible objective is to minimize the trace distance between $K_N$ and a target quantum gate $K_T$:
\begin{eqnarray}
F_0 = 1 - |\Tr(K^{\dagger}_TK_N)/D|^2,
\end{eqnarray}
where $D$ is the Hilbert space dimension.
The complete QOC formulation
includes secondary objectives and additional constraints
One way to address this formulation is by
adding to the objective
function weighted penalty terms representing constraint violation:
$ \min_{u_{k,j}} \left(\sum_{i=0}^{2}w_{i}F_{i} + \sum_{i=3}^{6}w_{i}G_{i}\right)$,
where

\begin{eqnarray}
F_1  &=& 1 - \frac{1}{n}\sum_j|\Tr(K^{\dagger}_TK_j)/D|^2 \\
F_2  &=&  1 - \frac{1}{n}\sum_j|\langle\psi_T|\psi_j\rangle|^2\\
G_3  &=&  1 - |\langle\psi_T|\psi_n\rangle|^2 \\
G_4  &=& |u|^2 \\
G_5  &=&  \sum_{k,j}|u_{j,k} - u_{k,j-1}|^2 \\
G_6  &=&  \sum_j|\langle\psi_F|\psi_j\rangle|^2
\end{eqnarray}
\noindent and $\psi_F$ is a forbidden state.

\begin{algorithm}
	\caption{Pseudocode for the GRAPE algorithm.}
	\label{algo:grape}
	\begin{algorithmic}
		\STATE Guess initial controls $u_{k,j}$.
		\REPEAT
		    \STATE Starting from $H_0$, calculate \\ { \quad\quad\quad $\rho_j=U_jU_{j-1}\ldots U_1 H_0 U_1^\dagger \ldots U_{j-1}^\dagger U_{j}^\dagger$}.
		    \STATE Starting from $\lambda_N =K_T$, calculate \\ { \quad\quad\quad $\lambda_j=U_{j+1}^\dagger \ldots U_N^\dagger K_T U_N \ldots U_{j}$}.
		    \STATE Evaluate $\frac{ \partial \rho_j \lambda_j }{\partial u_{k,j}}$
	            \STATE Update the $m \times N$ control amplitudes: \\ \quad\quad\quad $u_{j,k} \rightarrow u_{j,k}+\epsilon \frac{ \partial \rho_j \lambda_j }{\partial u_{k,j}}$
		\UNTIL{ $\Tr{(K_T^{\dagger}K_N)} <$ threshold}
		\STATE \textbf{return} $u_{j,k}$
	\end{algorithmic}
\end{algorithm}

QOC can be solved by several algorithms, including the gradient ascent pulse engineering (GRAPE) algorithm~\cite{PhysRevA.63.032308}. A basic version of GRAPE is shown in Algorithm~\ref{algo:grape}.
The derivatives $\frac{ \partial \rho_j \lambda_j }{\partial u_{k,j}}$ required by GRAPE can be calculated by hand coding or finite differences. Recently, these values have been calculated efficiently by automatic differentiation (AD or autodiff)~\cite{PhysRevA.95.042318}.

AD is a technique for transforming
algorithms that compute some mathematical function into algorithms
that compute the derivatives of that function~\cite{Griewank2008EDP,Naumann_book,baydin2015automatic}. AD works by differentiating the functions intrinsic to a given programming
language ({\tt sin()}, {\tt cos()}, {\tt +}, {\tt -}, etc.) and combining the partial derivatives using the chain rule of differential calculus.
The associativity of the chain rule leads to
two main methods of combining partial derivatives.  The forward mode
combines partial derivatives
starting with the independent variables and propagating forward to the
dependent variables. The reverse mode combines
partial derivatives starting with the dependent variables and
propagating back to the independent variables.
It is particularly attractive in the case of scalar functions, where a
gradient of arbitrary length can be computed at a fixed multiple of the operations count of the function.

\begin{figure}[htbp]
\vspace{-2em}
\centering {\includegraphics[width=0.8\textwidth]{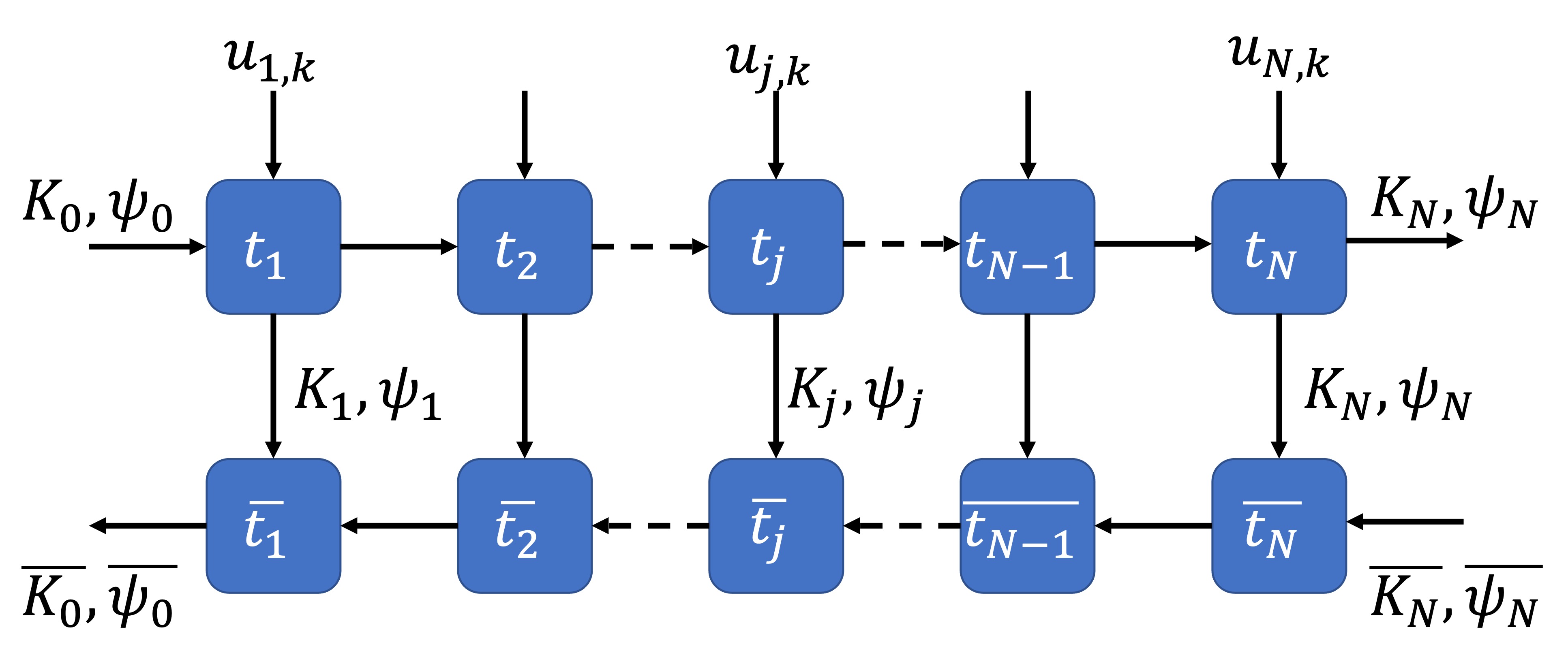}}
\vspace{-1em}
\caption{Reverse-mode gradient computation for QOC. Each forward time step, $j$, computes $K_j$,
$U_j$, and $\psi_j$ and stores them in memory. The reverse sweep starts at time step $N$. Each reverse time step $j$ uses the previously stored $K_j$,
$U_j$, and $\psi_j$.}
\label{fig:derivative}
\end{figure}

The reverse mode of AD is appropriate for QOC because of
the large number ($m \times N$) of inputs and the small
number of outputs (the cost function(s)).
As shown in Figure~\ref{fig:derivative}, standard reverse-mode
AD stores the results of intermediate time steps ($K_j$,
$U_j$, $\psi_j$) in order to compute $\frac{ \partial
\rho_j \lambda_j }{\partial u_{k,j}}$. This implies
that reverse-mode AD requires additional memory that is exponentially
proportional to $q$. Current QOC simulations therefore are limited in both the number of qubits that can be simulated and the number of time steps in the simulation. In this work we explore the suitability of {\em checkpointing} as well as {\em unitary matrix reversibility} to overcome this additional memory requirement.

The rest of the paper is organized as follows. Section~\ref{sec:related} discusses related work.
Section~\ref{sec:memoryrequirements} presents our approach to reducing the memory requirements of QOC.
Section~\ref{sec:impl} details the QOC implementation in JAX, and the evaluation of this approach is presented in Section~\ref{sec:results}. Section~\ref{sec:conclusion} concludes the paper and discusses future work.

\section{Related Work}
\label{sec:related}
QOC has been implemented in several packages, such as the Quantum Toolbox in Python (QuTIP) ~\cite{JOHANSSON20121760,JOHANSSON20131234}.
In addition to GRAPE, QOC can  be solved by  using the  chopped random basis (CRAB) algorithm~\cite{PhysRevLett.106.190501,PhysRevA.84.022326}. The problem is
formulated as the extremization of a multivariable function, which can be numerically approached with a suitable method such as steepest descent or conjugate gradient.
If computing the gradient is expensive, CRAB can instead use a derivative-free optimization algorithm. In~\cite{PhysRevA.95.042318}, the AD capabilities of TensorFlow are used to compute gradients for QOC.

Checkpointing is a well-established approach in AD to reduce the memory requirements of reverse-mode AD~\cite{griewank1992achieving,Recursive}.  In short, checkpointing techniques trade recomputation for storing intermediate states; see Section~\ref{sec:checkpointing} for more details.  For time-stepping codes, such as QOC, checkpointing strategies can range from simple periodic schemes~\cite{Griewank2008EDP}, through binomial checkpointing schemes~\cite{10.1145/347837.347846} that minimize recomputation subject to memory constraints, to multilevel checkpointing schemes~\cite{10.1016/j.procs.2016.05.444,aupy2016optimal} that store checkpoints to a multilevel storage hierarchy. Checkpointing schemes have also been adapted to deep neural networks~\cite{chen2016training,jain2020checkmate,10.1145/3458817.3476205,beaumont2020optimal} and combined with checkpoint compression~\cite{cyr2015towards,kukreja2020lossy}.


\section{Reducing Memory Requirements}
\label{sec:memoryrequirements}
We explore three
approaches to
reduce the
memory required to compute the derivatives for QOC.

\subsection{Approach 1: Checkpointing}
\label{sec:checkpointing}
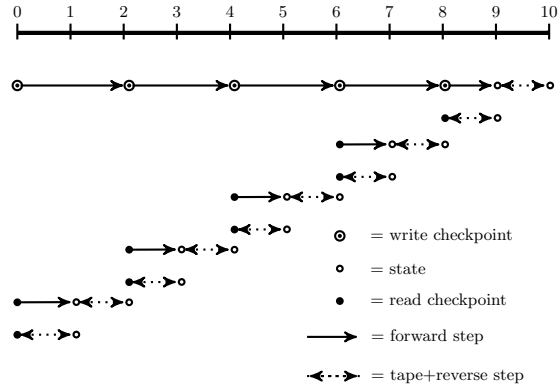
\begin{figure}[htb]
  \begin{center}
  \begin{tikzpicture}[>=stealth',shorten >=1pt,thick,black,auto, scale = 0.7, transform shape,initial/.style={black,fill=black,text=white}]
\tikzset{every state/.style={minimum size=0pt, minimum width=0pt, minimum height=0pt,scale=.35,black}}
  \foreach \x in {0,1,2,...,10}
    {
      \coordinate (A\x) at ($(0,0)+(\x*1.01cm,0)$) {};
      \draw ($(A\x)+(0,5pt)$) -- ($(A\x)-(0,5pt)$);
      \node at ($(A\x)+(0,3ex)$) {\x};
    }
  \draw[ultra thick] (A0) -- (A10) -- ($(A10)+0.01*(1,0)$);

\node[state,accepting]         (A)    at (0,-1)                 {};
\node[state,white]             (B) [right=1cm of A]             {};
\node[state,accepting]         (C) [right=2cm of A]             {};
\node[state,white]             (D) [right=3cm of A]             {};
\node[state,accepting]         (E) [right=4cm of A]             {};
\node[state,white]             (F) [right=5cm of A]             {};
\node[state,accepting]         (G) [right=6cm of A]             {};
\node[state,white]             (H) [right=7cm of A]             {};
\node[state,accepting]         (I) [right=8cm of A]             {};
\node[state]                   (J) [right=9cm of A]             {};
\node[state]                   (K) [right=10cm of A]            {};

\node[state,fill=black]        (I1) [below=0.5cm of I]           {};
\node[state]                   (J1) [below=0.5cm of J]            {};

\node[state,fill=black]        (G2) [below=1cm of G]            {};
\node[state]                   (H2) [below=1cm of H]            {};
\node[state]                   (I2) [below=1cm of I]            {};
\node[state,fill=black]        (G3) [below=0.5cm of G2]         {};
\node[state]                   (H3) [below=0.5cm of H2]         {};

\node[state,fill=black]        (E4) [below=2cm of E]            {};
\node[state]                   (F4) [below=2cm of F]            {};
\node[state]                   (G4) [below=2cm of G]            {};
\node[state,fill=black]        (E5) [below=0.5cm of E4]         {};
\node[state]                   (F5) [below=0.5cm of F4]         {};

\node[state,fill=black]        (C6) [below=3cm of C]            {};
\node[state]                   (D6) [below=3cm of D]            {};
\node[state]                   (E6) [below=3cm of E]            {};
\node[state,fill=black]        (C7) [below=0.5cm of C6]         {};
\node[state]                   (D7) [below=0.5cm of D6]         {};

\node[state,fill=black]        (A8) [below=4cm of A]            {};
\node[state]                   (B8) [below=4cm of B]            {};
\node[state]                   (C8) [below=4cm of C]            {};
\node[state,fill=black]        (A9) [below=0.5cm of A8]         {};
\node[state]                   (B9) [below=0.5cm of B8]         {};

\node[state,accepting]         (L4) [below=1cm of G3,label={}]    {};
\node[]                        (L5) [right=0.4cm of L4]    {= write checkpoint};
\node[state]                   (L6) [below=0.5cm of L4,label={}]    {};
\node[]                          (L16) [right=0.4cm of L6]    {= state};
\node[state,fill=black]   (L7) [below=0.5cm of L6,label={}]    {};
\node[]                        (L8) [right=0.4cm of L7]    {= read checkpoint};
\node[]                        (L9) [below=0.5cm of L7,label={}]    {};
\node[]                        (L10) [left=0.5cm of L9,label={}]    {};
\node[]                        (L11) [right=1cm of L10,label={}]    {};
\node[]                        (L12) [right=-0.15cm of L11]    {= forward step};
\node[]                        (L13) [below=0.5cm of L10,label={}]    {};
\node[]                        (L14) [below=0.5cm of L11,label={}]    {};
\node[]                        (L15) [right=-0.15cm of L14]    {= tape+reverse step};

\path[->] (A) edge node [right] {}  (C)
       (C) edge node [right] {}  (E)
       (E) edge node [right] {}  (G)
       (G) edge node [right] {}  (I);
\path[->] (I) edge node [right] {}  (J);
\path[<->] (J) edge[dotted] node [right] {}  (K);
\path[<->] (I1) edge[dotted] node [right] {}  (J1);

\path[->] (G2) edge node [right] {}  (H2);
\path[<->] (H2) edge[dotted] node [right] {}  (I2);
\path[<->] (G3) edge[dotted] node [right] {}  (H3);

\path[->] (E4) edge node [right] {}  (F4);
\path[<->] (F4) edge[dotted] node [right] {}  (G4);
\path[<->] (E5) edge[dotted] node [right] {}  (F5);

\path[->] (C6) edge node [right] {}  (D6);
\path[<->] (D6) edge[dotted] node [right] {}  (E6);
\path[<->] (C7) edge[dotted] node [right] {}  (D7);

\path[->] (A8) edge node [right] {}  (B8);
\path[<->] (B8) edge[dotted] node [right] {}  (C8);
\path[<->] (A9) edge[dotted] node [right] {}  (B9);

\path[->] (L10) edge node [right] {}  (L11);
\path[->] (L13) edge[dotted] node [right] {}  (L14);
\path[->] (L14) edge[dotted] node [right] {}  (L13);

\end{tikzpicture}
   \vspace{-1em}
  \end{center}
\caption{Periodic checkpointing schedule for $N\!=\!10$ time steps and $5$ checkpoints ($C\!=\!2$).}
	\label{fig:periodiccheckpointing}
\end{figure}

Checkpointing schemes reduce the memory requirements of reverse-mode AD by recomputing certain intermediate states instead of storing them.
These schemes  checkpoint the inputs of selected time steps in a {\em plain-forward} sweep. To compute the gradient, a stored checkpoint is read, followed by a forward sweep and a reverse sweep for an appropriate number of time steps. Figure~\ref{fig:periodiccheckpointing} illustrates periodic checkpointing for a computation of $10$ time steps and $5$ checkpoints. In the case of QOC with $N$ time steps and periodic checkpointing interval $C$, one must store $O(C + \frac{N}{C})$  matrices of size $2^q\times2^q$.

\subsection{Approach 2: Reversibility of Unitary Matrices}
The second approach is to
exploit the property of unitary matrices that the inverse of a unitary matrices is its conjugate transpose.
\begin{eqnarray}
U^\dagger U &=& UU^\dagger \\
U^\dagger &=& U^{-1}
\label{eq:inverse}
\end{eqnarray}
Computing the
inverse by exploiting the reversibility property of unitary matrices is an exact and inexpensive process.
The use of the inverse allows us to compute $K_{j-1}$ from $K_{j}$ and $\psi_{j-1}$ from $\psi_{j}$.
\begin{eqnarray}
K_j & = & U_{j}U_{j-1}U_{j-2} \ldots U_{1}U_{0}\\
\label{eq:useinverse}
K_{j-1}& = & U_{j}^\dagger K_j\\
\label{eq:useinversestate}
\psi_{j-1}&=&\psi_0 K_{j-1}
\end{eqnarray}
Thus, one does not have to store any of the $K_j$ matrices required to compute the adjoint of
a time step. This approach reduces the memory requirement by half. 

More importantly, using reversibility can unlock a further reduction in
the memory requirements by not storing $U_j$,
 but rather using only the $u_{j,k}$ control values to recompute $U_j$. As a result, no intermediate computations need to be stored, and therefore the only additional requirements are to store the derivatives of the function with respect to the controls and other variables, thus basically doubling the memory requirements relative to the function itself.

\subsection{Approach 3: Periodic Checkpointing Plus Reversibility}

The reversibility property of unitary matrices is exact only in real arithmetic. A floating-point implementation may incur roundoff errors.  Therefore, Equation~\ref{eq:useinverse} might not hold exactly, especially for large numbers of time steps.  That is,
\begin{eqnarray}
K_{j-1}&\approx& U_{j}^\dagger K_j\\
\psi_{j-1}&\approx&\psi_0 K_{j-1}\label{eq:useinverse1}
\end{eqnarray}
in floating-point arithmetic. Because $K_{j-1}$ is computed each time step, the error continues to grow as the computation proceeds in the reverse sweep.

To mitigate this effect, we can combine the two approaches, checkpointing every $C$ time steps and, during the reverse pass,
instead of computing forward from these checkpoints, computing backward from the checkpoints by exploiting reversibility. Thus, floating-point errors in Equation~\ref{eq:useinverse} are incurred over a maximum of $C$ time steps, and we reduce the number of matrices of size $2^q\times2^q$ stored from $O(C + \frac{N}{C})$ to $O(\frac{N}{C})$.


\begin{table}[]
\caption{Overview of the object sizes,  number of object instances stored for the forward computation, and number of additional instances that need to be stored for store-all, checkpointing, reversibility, and checkpointing plus reversibility. The total memory size in the last row is the product of the object size and the number of instances.
}
\label{tab:memory}
\bgroup
\def\arraystretch{1.1}
\setlength\tabcolsep{4.3pt}
\begin{tabular}{l|l|l|l|l|l|l}
\textbf{Variable} & \textbf{Size} & \textbf{Forward} & \textbf{Store} & \textbf{Checkpoint} & \textbf{Revert} & \textbf{Rev + Ckp} \\ \hline
$u_{k,j}$ & $1$ & $Nm$ &$+0$ & $+0$ & $+0$ & $+0$ \\
$H$ & $2^q \cdot 2^q$ & $m$ & $+0$ & $+0$ & $+0$ & $+0$ \\
$\mathbb{H}_j$ & $2^q \cdot 2^q$ & $1$ & $+0$ & $+0$ & $+0$ & $+0$ \\
$U_j$ & $2^q \cdot 2^q$ & $1$ & $+N$ & $+C$ & $+0$ & $+0$ \\
$K_j$ & $2^q \cdot 2^q$ & $1$ & $+N$ & $+\frac{N}{C} + C$ & $+0$ & $+\frac{N}{C}$ \\
$\psi_j$ & $2^q$ & $1$ & $+N$ & $+\frac{N}{C} + C$ & $+0$ & $+\frac{N}{C}$ \\ \hline
\textbf{Mem ($\mathcal{O}$)} & & $2^{2q}m+ Nm$ & $+{2^{2q}N}$ & $+{2^{2q}\left(\frac{N}{C}+ C\right)}$ & $+{0}$ & $+{2^{2q}\frac{N}{C}}$
\end{tabular}
\egroup
\end{table}

\subsection{Analysis of memory requirements}

Table~\ref{tab:memory} summarizes the memory requirements for the forward pass of the function evaluation as well as the added cost of the various strategies for computing the gradient.  Conventional AD, which stores the intermediate states $U_j$, $K_j$, and $\psi_j$ at every time step, incurs an additional storage cost proportional to the number of time steps \textit{times} the size of the $2^q \times 2^q$ matrices.  Periodic checkpointing reduces the number of matrices stored to $\frac{N}{C}+C$. The checkpointing interval that minimizes this cost occurs when $\frac{\partial}{\partial C} \left(\frac{N}{C}+C\right) = \frac{-N}{C^2} + 1 = 0$, or $C = \sqrt{N}$. Exploiting reversibility enables one to compute $U_j$ from $u_{j,k}$ and $H_k$ and $K_{j-1}$ from $K_{j}$ and $U_{j}$, resulting in essentially zero additional memory requirements, beyond those required to store the derivatives themselves.  Combining reversibility with periodic checkpointing eliminates the number of copies of $U_j$ and $K_j$ to be stored from $\frac{N}{C} + C$ to $\frac{N}{C}$.

\section{Implementation}
\label{sec:impl}
As an initial step we have ported to the JAX machine learning framework~\cite{jax2018github} a version of QOC that was
previously implemented in TensorFlow.
JAX provides a NumPy-style interface and supports execution on CPU systems as well as GPU and TPU (tensor processing unit) accelerators, with built-in automatic differentiation and just-in-time compilation capability. JAX supports checkpointing through the use of the {\tt jax.checkpoint} decorator and allows custom derivatives to be created for functions using the {\tt custom\_jvp} decorator for forward mode and the {\tt custom\_vjp} decorator for reverse mode. To enable our work, we have contributed {\tt jax.scipy.linalg.expm} to JAX to perform the matrix exponentiation operation using Pad\'e approximation and to compute its derivatives. This code is now part of standard JAX releases.

Our approach requires us to perform checkpointing or use custom derivatives only for the Python function that implements Equations~\ref{evovlveschrodingersicrete1}--\ref{evovlveschrodingersicrete4} for a single time step $j$ or a loop over a block of time  steps.
Standard AD can be used as before for the rest of the code. By implication, our approach does not change for different objective functions.

We show here the implementation of the periodic checkpointing plus reversibility approach
and direct the reader to our open source implementation for further details~\cite{qocjaxweb}. Listing~\ref{lst:evolve_step_loop} shows the primal code that computes a set of time steps.
The function \texttt{evolve\_step} computes Equations~\ref{evovlveschrodingersicrete1}--\ref{evovlveschrodingersicrete4}.

\begin{lstfloat}
\begin{lstlisting}[style=Python]
def evolve_step_loop(start, stop, cost_eval_step, dt, states, K,
                     control_eval_times, controls):
    for step in range(start,stop):
        # Evolve the states and K to the next time step.
        time = step * dt
        states, K  = evolve_step(dt, states, K, time,
                                         control_eval_times, controls)
    return states, K
\end{lstlisting}
\caption{Simplified code showing a loop to simulate QOC for N time steps}
\label{lst:evolve_step_loop}
\end{lstfloat}

\begin{lstfloat}
\begin{lstlisting}[style=Python]
@jax.custom_vjp
def evolve_step_loop_custom(start, stop, cost_eval_step, dt, states,
                            K, control_eval_times, controls):
    states, K = evolve_step_loop(start, stop, cost_eval_step, dt,
                                         states,K,control_eval_times,
                                         controls)
    return states, K
\end{lstlisting}
\caption{A wrapper to {\tt evolve\_step\_loop()}, which will have  derivatives provided by the user.}
\label{lst:evolve_step_custom}
\end{lstfloat}

Listing~\ref{lst:evolve_step_custom} is a convenience wrapper with for the primal code. We decorate the wrapper with {\tt \@jax.custom\_vjp} to inform JAX that we will provide custom derivatives for it. User-provided custom derivatives for a JAX function consist of a forward sweep and a reverse sweep.  The forward sweep must store all the information required to compute the derivatives in the reverse sweep. Listing~\ref{lst:evolve_step_custom_fwd} is the provided forward sweep. Here, as indicated in Table~\ref{tab:memory}, we are storing the $K$ matrix and the state vector. Note that this form of storage is effectively a checkpoint, even though it does not use {\tt jax.checkpoint}.

\begin{lstfloat}
\begin{lstlisting}[style=Python]
def evolve_loop_custom_fwd(start, stop,cost_eval_step, dt, states, K,
                           control_eval_times, controls):
    states, K = _evaluate_schroedinger_discrete_loop_inner(
                           start, stop, cost_eval_step, dt, states, K,
                           control_eval_times, controls)
    #Here we store the final state and K for use in the backward pass
    return (states,K), (start, stop, cost_eval_step, dt, states,
           K, control_eval_times,controls)

\end{lstlisting}
\caption{Forward sweep of the user-provided derivatives.}
\label{lst:evolve_step_custom_fwd}
\end{lstfloat}

 Listing~\ref{lst:evolve_step_custom_bwd} is a user-provided reverse sweep. It starts by restoring the values passed to it by the forward sweep.
 While looping over time steps in reverse order, it recomputes Equations~\ref{evovlveschrodingersicrete1}--\ref{evovlveschrodingersicrete2}. It then computes Equations ~\ref{eq:inverse}, ~\ref{eq:useinverse}, and ~\ref{eq:useinversestate} to retrieve
$K_{j-1}$ and $\psi_{j-1}$, which are then used to compute the adjoints for the time step. The code to compute the adjoint of the time step was obtained by the source transformation AD tool Tapenade~\cite{hascoet2013tapenade}.
\begin{lstfloat}
\begin{lstlisting}[style=Python]
def evolve_loop_custom_bwd(res,g_prod):
  #Restore all the values stored in the forward sweep
  start, stop, cost_eval_step, dt, states,
        K, control_eval_times, controls = res
  _M2_C1 = 0.5
  controlsb = jnp.zeros(controls.shape, states.dtype)
  #Go backwards in time steps
  for i in range(stop-1,start-1,-1):
    #Reapply controls to compute a step unitary matrix
    time = i * dt
    t1 = time + dt * _M2_C1
    x = t1
    xs = control_eval_times
    ys = controls
    index = jnp.argmax(x <= xs)
    y = ys[index - 1] + (((ys[index] - ys[index - 1]) /
        (xs[index] - xs[index - 1])) * (x - xs[index - 1]))
    controls_ = y
    hamiltonian_ = (SYSTEM_HAMILTONIAN
                + controls_[0] * CONTROL_0
                + jnp.conjugate(controls_[0]) * CONTROL_0_DAGGER
                + controls_[1] * CONTROL_1
                + jnp.conjugate(controls_[1]) * CONTROL_1_DAGGER)
    a1 = -1j * hamiltonian_
    magnus = dt * a1
    step_unitary, f_expm_grad = jax.vjp(jax.scipy.linalg.expm, (magnus),
    has_aux=False)
    #Exploit reversibility of unitary matrix
    #and calculate previous state and K
    step_unitary_inv=jnp.conj(jnp.transpose(step_unitary))
    states=(jnp.matmul(step_unitary_inv,states))
    K=(jnp.matmul(step_unitary_inv,K))
    _, f_matmul = jax.vjp(jnp.matmul,step_unitary, states)
    _, f_matmul_K = jax.vjp(jnp.matmul,step_unitary, K)
    #Go backwards for the timestep
    step_unitaryb,Kb=f_matmul_K(g_prod[1])
    step_unitaryb,statesb=f_matmul(g_prod[0])
    magnusb = f_expm_grad(step_unitaryb)
    a1b=dt*magnusb[0]
    hamiltonian_b = jnp.conjugate(-1j)*a1b
    controls1b=jnp.array((jnp.sum(jnp.conjugate(CONTROL_0)*hamiltonian_b) +
      jnp.conjugate(jnp.sum(jnp.conjugate(CONTROL_0_DAGGER)*hamiltonian_b)),
      jnp.sum(jnp.conjugate(CONTROL_1)*hamiltonian_b) +
      jnp.conjugate(jnp.sum(jnp.conjugate(CONTROL_1_DAGGER)*hamiltonian_b))),
      dtype=hamiltonian_b.dtype)
    tempb = (x-control_eval_times[index-1])*controls1b/
      (control_eval_times[index]-control_eval_times[index-1])
    controlsb=jax.ops.index_update(controlsb,
      jax.ops.index[index-1],controlsb[index-1]+controls1b - tempb)
    controlsb=jax.ops.index_update(controlsb,
      jax.ops.index[index],controlsb[index]+tempb)
    g_prod=statesb,Kb
  return (0.0,0.0,0.0,0.0,statesb,Kb,0.0,-1*controlsb)

  evolve_loop_custom.defvjp(evolve_loop_custom_fwd, evolve_loop_custom_bwd)
\end{lstlisting}
\caption{User-provided reverse sweep that exploits reversibility. }
\label{lst:evolve_step_custom_bwd}
\end{lstfloat}

\section{Experimental Results}
\label{sec:results}
We compared standard AD, periodic checkpointing, and full reversibility or periodic checkpointing with reversibility, as appropriate.
We conducted our experiments on a cluster where each compute node was connected to 8 NVIDIA A100 40GB GPUs. Each node contained
1TB DDR4 memory and 320GB GPU memory. We validated the output of the checkpointing and reversibility approaches against the standard approach implemented using JAX.  We used the JAX memory profiling capability in conjunction with {\tt GO pprof} to measure the memory needs for each case. We conducted three sets of experiments to evaluate the approaches, varying the number of qubits, the number of time steps, or the checkpoint period.

\begin{figure} [htbp]
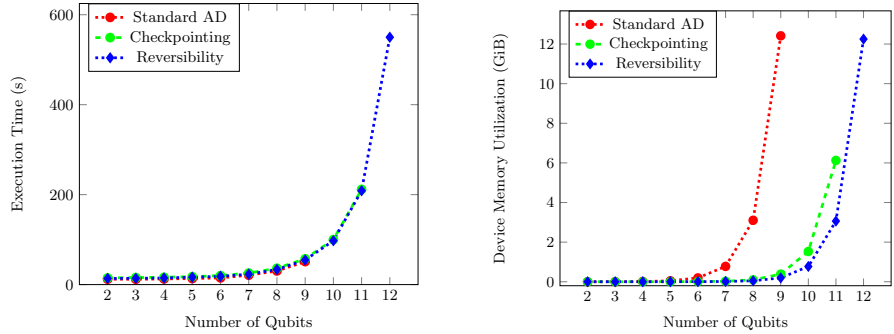

\begin{center}
    \includestandalone[width=.45\textwidth]{figures/vary_qubits_time}\hspace{1.5em}
    \includestandalone[width=.44\textwidth]{figures/vary_qubits_mem}
    \end{center} \vspace{-1em}
    \caption{Comparison of execution time and device memory requirements for standard AD, periodic checkpointing, and full reversibility with increasing number of qubits. The QOC simulation consisted of 100 time steps with a checkpoint period of 10. \label{fig:vary_qubits}}
\end{figure}

\subsection{Vary Qubits}
We first varied the number of qubits $q$, keeping the number of time steps fixed at $100$ and the checkpoint period fixed at $C=\sqrt{N}=10$.  Figure~\ref{fig:vary_qubits}
shows the memory consumed by standard AD, periodic checkpointing, and full reversibility.
One can see that the device memory requirements for the standard approach are highest whereas the requirements for reversibility are lowest, although all three grow exponentially as a function of $q$, as predicted by the analysis in Section~\ref{sec:memoryrequirements}.
Furthermore, we note that the standard approach can be executed for a maximum of $9$ qubits and runs out of available device memory on the $10$th qubit. The periodic checkpointing approach can be run for $11$ qubits and runs out of available device memory on the $12$th. The full reversibility approach can be run for $12$ qubits and exceeds available device memory on the $13$th.  Figure~\ref{fig:vary_qubits} (left) also shows the execution time for the various approaches. The times are similar for the cases that can be executed before running out of memory.  As expected, the time grows exponentially as a function of $q$.

\begin{figure} [htbp]
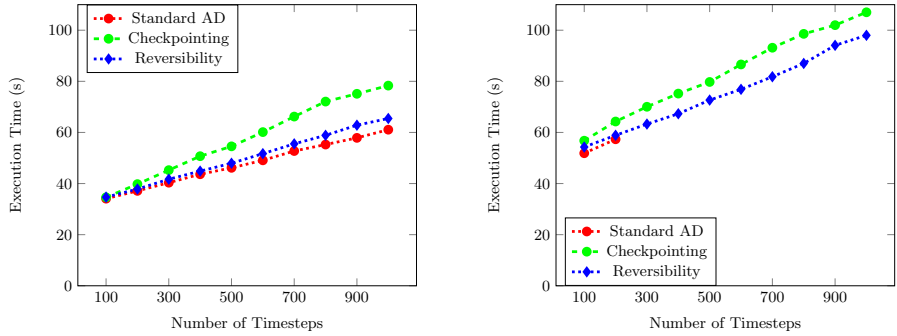

\begin{center}
    \includestandalone[width=.45\textwidth]{figures/vary_steps_time_8} \hspace{1em}
\includestandalone[width=.45\textwidth]{figures/vary_steps_time_9}
\end{center}
    \caption{Comparison of the execution time for  standard AD, periodic checkpointing, and periodic reversibility approaches with increasing number of time steps. The QOC simulation consisted of $8$ (left) or $9$ (right) qubits. The checkpoint period was chosen to be the square root of the number of time steps. \label{fig:vary_steps_time}}
\end{figure}

\begin{figure} [htbp]
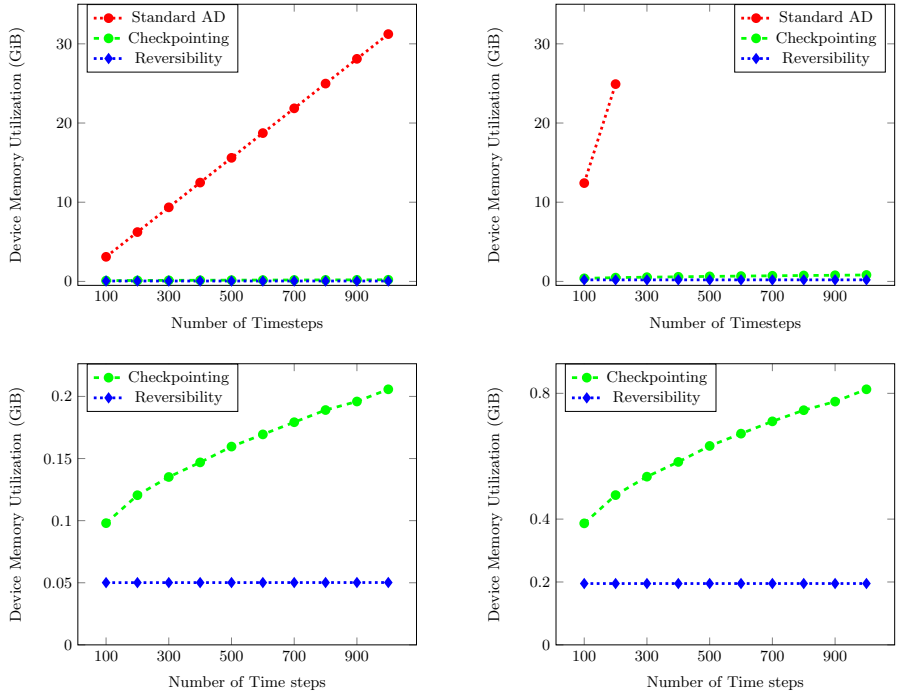

\begin{center}
        \includestandalone[width=.45\textwidth]{figures/vary_steps_mem_8} \hspace{1em}
    \includestandalone[width=.45\textwidth]{figures/vary_steps_mem_9}\\[1em]
        \includestandalone[width=.45\textwidth]{figures/vary_steps_check_rev_mem_8} \hspace{1em}
    \includestandalone[width=.45\textwidth]{figures/vary_steps_check_rev_mem_9}
\end{center}
    \caption{Comparison of the device memory requirements for  standard AD, periodic checkpointing, and periodic reversibility approaches with increasing number of time steps. The QOC simulation consisted of $8$ (left) or $9$ (right) qubits. The checkpoint period was chosen to be the square root of the number of time steps. Top row shows all three approaches; bottom row omits standard AD. \label{fig:vary_steps_mem}}
\end{figure}

\subsection{Vary Time Steps}
Next we fixed the number of qubits at $q=8$ or $q=9$ and varied the number of time steps, $N$. For periodic checkpointing we used the optimal checkpoint period, $C=\sqrt{N}$.  We expect
the time to be roughly linear in $N$ and independent of $C$ because every $U_j$ and $K_j$ must be computed once during the forward pass and one more time on the reverse pass.  We expect periodic checkpointing and full reversibility to be slower than standard AD because they both trade some amount of recomputation for reduced storage requirements.  We expect full reversibility to be somewhat faster than periodic checkpointing alone because periodic checkpointing must compute forward from the checkpoint, storing intermediate $K_j$ along the way, while full reversibility skips the second forward pass and is able to restore $K_j$ during the reverse pass directly from the controls and $K_{j+1}$. Figure~\ref{fig:vary_steps_time} is consistent with these expectations, although standard AD quickly runs out of memory for the case $q=9$.

Based on the analysis in Section~\ref{sec:memoryrequirements}, we expect the memory requirements of standard AD to be linear in the number of time steps and the memory requirements of full reversibility to be independent of the number of time steps. We expect the memory requirements of periodic checkpointing to vary as a function of $\frac{N}{C}+C$; since $C$ is chosen to be $C=\sqrt{N}$, the memory should vary as a function of $\sqrt{N}$. Figure~\ref{fig:vary_steps_mem} clearly shows the linear dependence of standard AD and independence of full reversibility on the number of time steps. The memory requirements for periodic checkpointing are also consistent with expectations.

\begin{figure} [t!]
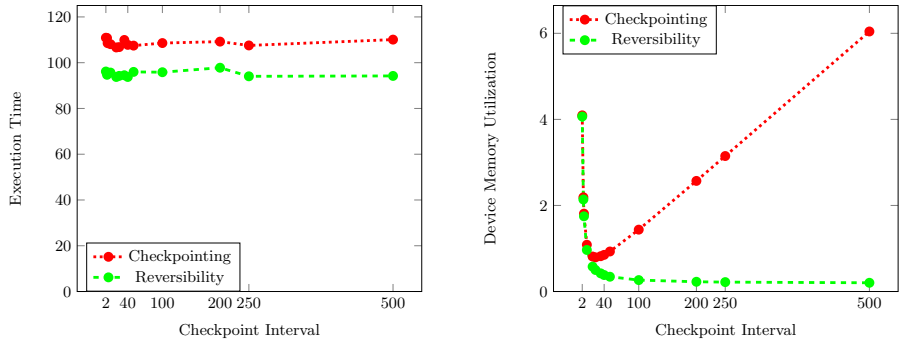

\begin{center}
    \includestandalone[height=4.5cm]{figures/vary_checks_time}\hspace{1em}
    \includestandalone[height=4.5cm]{figures/vary_checks_mem}
\end{center}
    \caption{Comparison of the execution time and device memory requirements for  periodic checkpointing and checkpointing plus reversibility approaches with increasing number of time steps. The QOC simulation consisted of $1,000$ time steps and $9$ qubits.
    }
    \label{fig:vary_checks}
\end{figure}
\subsection{Vary Checkpointing Period}

We examined the dependence of execution time and memory requirements on the checkpointing period, $C$, keeping the number of qubits fixed at $q=9$ and the number of time steps fixed at $N=1,000$. We expect the time to be roughly independent of $C$ because every $U_j$ and $K_j$ must be computed once during the forward pass and one more time on the reverse pass.  We expect periodic checkpointing with reversibility to be somewhat faster than periodic checkpointing alone because periodic checkpointing must compute forward from the checkpoint, storing intermediate $K_j$ along the way, while periodic checkpointing with reversibility skips the second forward pass and is able to restore $K_j$ during the reverse pass directly from the controls and $K_{j+1}$. The timing results in Figure~\ref{fig:vary_checks} (left) are consistent with these expectations.

We expect the memory requirements of periodic checkpointing with reversibility to vary as a function of $\frac{N}{C}$ or, since $N$ is constant, as a function of $\frac{1}{C}$.  We expect the memory requirements of periodic checkpointing alone to vary as a function of $\frac{N}{C}+C$, with a minimum at $C=\sqrt{N} \approx 32$.  Again, the memory utilization results in Figure~\ref{fig:vary_checks} (right) are consistent with these expectations.

\section{Conclusion and Future Work}
\label{sec:conclusion}
We have implemented a version of quantum optimal control (QOC) using the JAX framework. We have compared standard automatic differentiation (AD), periodic checkpointing, and reversibility---a nonstandard AD approach
that recognizes that the inverse of a unitary matrix is its conjugate transpose. Checkpointing and reversibility are both superior to standard AD. The reversibility approach, however, allows  more qubits to be simulated when the number of time steps is large.
%
%
%
Recognizing that reversibility
(Equation~\ref{eq:useinverse})
is precise in real arithmetic but is not
precise in floating-point arithmetic, we demonstrated that reversibility can be combined with  periodic checkpointing, reducing memory requirements relative to periodic checkpointing alone while ensuring that roundoff errors are not accumulated over a period of more than $C$ time steps.

In the future, we will study methods to estimate the amount of roundoff error as a function of $C$ in order to choose a period that minimizes memory requirements while incurring acceptable roundoff errors.  We will investigate applying lossy compression to the checkpoints and compare the trade-offs in storage and accuracy between periodic checkpointing with lossy compression and periodic checkpointing with reversibility.  Moreover, we will combine periodic checkpointing, lossy compression, and reversibility to enable QOC to be applied to even larger numbers of qubits and time steps.

\bibliographystyle{splncs04}
\bibliography{main}

\vspace{-7.3in}\hspace{-1.2in}
\scriptsize
\rotatebox{90}{
\framebox{\parbox{7in}{The submitted manuscript has been created by UChicago Argonne, LLC, Operator of Argonne National Laboratory (`Argonne'). Argonne, a U.S. Department of Energy Office of Science laboratory, is operated under Contract No. DE-AC02-06CH11357. The U.S. Government retains for itself, and others acting on its behalf, a paid-up nonexclusive, irrevocable worldwide license in said article to reproduce, prepare derivative works, distribute copies to the public, and perform publicly and display publicly, by or on behalf of the Government.  The Department of Energy will provide public access to these results of federally sponsored research in accordance with the DOE Public Access Plan. \url{http://energy.gov/downloads/doe-public-access-plan}.}}
}
\normalsize

\end{document}

%% file: figures/vary_qubits_time.tex
\begin{tikzpicture}
   \begin{axis}[
                xlabel = {Number of Qubits},
                ylabel = { Execution Time (s)},
                xtick={2,3,4,5,6,7,8,9,10,11,12},
                ymin=0,
                ymax=625,
                mark options={solid},
                legend style={
                    at={(0.25,1.0)},
                    anchor=north,
                    legend columns = 1}
                    ]
      \addplot[ultra thick, red, dotted, mark=*] coordinates {
      (2 , 11.9956)
      (3 , 11.8696)
      (4 , 12.5923)
      (5 , 13.4402)
      (6 , 14.8775)
      (7 , 20.9349)
      (8 , 30.3309)
      (9 , 51.1096)
      };
      \addplot[ultra thick, green, dashed, mark=*] coordinates {
      (2 , 14.7865)
      (3 , 15.6192)
      (4 , 16.2456)
      (5 , 17.6264)
      (6 , 20.0220)
      (7 , 25.6519)
      (8 , 36.2769)
      (9 , 57.2887)
      (10 , 100.0035)
      (11 , 211.6795)
      };
      \addplot[ultra thick, blue, dotted, mark=diamond*] coordinates {
      (2 , 13.6033)
      (3 , 13.9597)
      (4 , 14.6043)
      (5 , 16.4639)
      (6 , 18.3222)
      (7 , 23.0783)
      (8 , 33.6712)
      (9 , 54.6943)
      (10 , 97.6287)
      (11 , 208.6915)
      (12 , 549.9379)
      };
      \legend{{Standard AD},
              {Checkpointing},
              {Reversibility}
        }
   \end{axis}
\end{tikzpicture}

%% file: figures/vary_qubits_mem.tex
\begin{tikzpicture}
   \begin{axis}[
                xlabel = {Number of Qubits},
                ylabel = {Device Memory Utilization (GiB)},
                xtick={2,3,4,5,6,7,8,9,10,11,12},
                ymin=-0.2,
                mark options={solid},
                legend style={
                    at={(0.25,1.0)},
                    anchor=north,
                    legend columns=1}
                    ]
      \addplot[ultra thick, red, dotted, mark=*] coordinates {
      (2, 0.001470131)
      (3, 0.003811359)
      (4, 0.012962074)
      (5, 0.049402271)
      (6, 0.195078765)
      (7, 0.775951732)
      (8, 3.103635658)
      (9, 12.41427695)
      };
      \addplot[ultra thick, green, dashed, mark=*] coordinates {
      (2, 0.000955667)
      (3, 0.001053553)
      (4, 0.001394548)
      (5, 0.00261754)
      (6, 0.007257675)
      (7, 0.025514279)
      (8, 0.097989385)
      (9, 0.386477393)
      (10, 1.531271499)
      (11, 6.125025313)
      };
      \addplot[ultra thick, blue, dotted, mark=diamond*] coordinates {
      (2, 0.00086631)
      (3, 0.000933827)
      (4, 0.001129703)
      (5, 0.001827803)
      (6, 0.004224651)
      (7, 0.013509088)
      (8, 0.050095522)
      (9, 0.195028366)
      (10, 0.771850825)
      (11, 3.062518466)
      (12, 12.2500261)
      };
      \legend{{Standard AD},
              {Checkpointing},
              {Reversibility}
        }
   \end{axis}
\end{tikzpicture}

%% file: figures/vary_steps_time_8.tex
\begin{tikzpicture}
   \begin{axis}[
                xlabel = {Number of Timesteps},
                ylabel = {Execution Time (s)}, 
                xtick={100,300,500,700,900},
                ymin=0,
                ymax=110,
                mark options={solid},
                legend style={
                    at={(0.25,1.0)},
                    anchor=north,
                    legend columns=1}
                    ]
      \addplot[ultra thick, red, dotted, mark=*] coordinates {
      (100 , 34.0869)
      (200 , 37.1298)
      (300 , 40.3354)
      (400 , 43.6368)
      (500 , 46.1479)
      (600 , 49.0895)
      (700 , 52.7292)
      (800 , 55.2636)
      (900 , 57.8748)
      (1000 , 61.0727)
      };
      \addplot[ultra thick, green, dashed, mark=*] coordinates {
      (100 , 34.6244) 
      (200 , 39.7843) 
      (300 , 45.2503) 
      (400 , 50.7038) 
      (500 , 54.5530) 
      (600 , 60.1336) 
      (700 , 66.2054) 
      (800 , 72.0795) 
      (900 , 75.1181) 
      (1000 , 78.3234)
      };
      \addplot[ultra thick, blue, dotted, mark=diamond*] coordinates {
      (100 , 34.6806)
      (200 , 37.8486)
      (300 , 41.6920)
      (400 , 44.8584)
      (500 , 47.9448)
      (600 , 51.7000)
      (700 , 55.5367)
      (800 , 58.8742)
      (900 , 62.8379)
      (1000 , 65.4575)
      };
      \legend{{Standard AD},
              {Checkpointing},
              {Reversibility}
        }
   \end{axis}
\end{tikzpicture}

%% file: figures/vary_steps_time_9.tex
\begin{tikzpicture}
   \begin{axis}[
                xlabel = {Number of Timesteps},
                ylabel = {Execution Time (s)}, 
                xtick={100,300,500,700,900},
                ymin=0,
                ymax=110,
                mark options={solid},
                legend style={
                    at={(0.25,0.0)},
                    anchor=south,
                    legend columns=1}
                    ]
      \addplot[ultra thick, red, dotted, mark=*] coordinates {
      (100 , 51.8646)
      (200 , 57.3265)
      };
      \addplot[ultra thick, green, dashed, mark=*] coordinates {
      (100 , 56.7678)
      (200 , 64.2784)
      (300 , 70.0430)
      (400 , 75.1990)
      (500 , 79.7517)
      (600 , 86.6023)
      (700 , 93.1451)
      (800 , 98.5709)
      (900 , 101.9772)
      (1000 , 107.0132)
      };
      \addplot[ultra thick, blue, dotted, mark=diamond*] coordinates {
      (100 , 54.2909)
      (200 , 58.8855)
      (300 , 63.2756)
      (400 , 67.3103)
      (500 , 72.6606)
      (600 , 76.8664)
      (700 , 81.7923)
      (800 , 86.9978)
      (900 , 94.0813)
      (1000 , 98.0023)
      };
      \legend{{Standard AD},
              {Checkpointing},
              {Reversibility}
        }
   \end{axis}
\end{tikzpicture}

%% file: figures/vary_steps_mem_8.tex
\begin{tikzpicture}
   \begin{axis}[
                xlabel = {Number of Timesteps},
                ylabel = {Device Memory Utilization (GiB)}, 
                xtick={100,300,500,700,900},
                ymin=-0.5,
                mark options={solid},
                ymax=35,
                ytick={0,10,20,30},
                yticklabels={\phantom{0.0}0,10,20,30},
                legend style={
                    at={(0.25,1.0)},
                    anchor=north,
                    legend columns=1}
                    ]
      \addplot[ultra thick, red, dotted, mark=*] coordinates {
      (100, 3.103635658)
      (200, 6.22875673)
      (300, 9.353877801)
      (400, 12.47899887)
      (500, 15.60411995)
      (600, 18.72924102)
      (700, 21.85436209)
      (800, 24.97948316)
      (900, 28.10460423)
      (1000, 31.2297253)
      };
      \addplot[ultra thick, green, dashed, mark=*] coordinates {
      (100, 0.097989385)
      (200, 0.120474321)
      (300, 0.135150102)
      (400, 0.146900897)
      (500, 0.15962888)
      (600, 0.16942991)
      (700, 0.179233802)
      (800, 0.18904115)
      (900, 0.195921545)
      (1000, 0.205734258)
      };
      \addplot[ultra thick, blue, dotted, mark=diamond*] coordinates {
      (100, 0.050095522)
      (200, 0.050106787)
      (300, 0.050117695)
      (400, 0.050128841)
      (500, 0.050139868)
      (600, 0.050150656)
      (700, 0.050161325)
      (800, 0.050171995)
      (900, 0.050182902)
      (1000, 0.050193452)
      };
      \legend{{Standard AD},
              {Checkpointing},
              {Reversibility}
        }
   \end{axis}
\end{tikzpicture}

%% file: figures/vary_steps_mem_9.tex
\begin{tikzpicture}
   \begin{axis}[
                xlabel = {Number of Timesteps},
                ylabel = {Device Memory Utilization (GiB)}, 
                xtick={100,300,500,700,900},
                ytick={0,10,20,30},
                yticklabels={\phantom{0.0}0,10,20,30},
                ymin=-0.5,
                mark options={solid},
                ymax=35,
                legend style={
                    at={(0.75,1.0)},
                    anchor=north,
                    legend columns=1}
                    ]
      \addplot[ultra thick, red, dotted, mark=*] coordinates {
      (100, 12.41427695)
      (200, 24.91449339)
      };
      \addplot[ultra thick, green, dashed, mark=*] coordinates {
      (100, 0.386477393)
      (200, 0.476345618)
      (300, 0.534966712)
      (400, 0.581873995)
      (500, 0.632687916)
      (600, 0.671785344)
      (700, 0.710885872)
      (800, 0.74634086)
      (900, 0.77372859)
      (1000, 0.812838059)
      };
      \addplot[ultra thick, blue, dotted, mark=diamond*] coordinates {
      (100, 0.195028366)
      (200, 0.195039631)
      (300, 0.195050539)
      (400, 0.195061685)
      (500, 0.195072711)
      (600, 0.1950835)
      (700, 0.195094169)
      (800, 0.195104838)
      (900, 0.195115746)
      (1000, 0.195126296)
      };
      \legend{{Standard AD},
              {Checkpointing},
              {Reversibility}
        }
   \end{axis}
\end{tikzpicture}

%% file: figures/vary_steps_check_rev_mem_8.tex
\begin{tikzpicture}
   \begin{axis}[
                xlabel = { Number of Time steps},
                ylabel = {Device Memory Utilization 
                (GiB)},  
                xtick={100,300,500,700,900},
                ytick={0,0.05,0.1,0.15,0.2,0.25},
                yticklabels={0,0.05,0.1,0.15,0.2,0.25},
                mark options={solid},
                ymin=0,
                legend style={
                    at={(0.25,1.0)},
                    anchor=north,
                    legend columns=1}
                    ]
      \addplot[ultra thick, green, dashed, mark=*] coordinates {
      (100, 0.097989385)
      (200, 0.120474321)
      (300, 0.135150102)
      (400, 0.146900897)
      (500, 0.15962888)
      (600, 0.16942991)
      (700, 0.179233802)
      (800, 0.18904115)
      (900, 0.195921545)
      (1000, 0.205734258)
      };
      \addplot[ultra thick, blue, dotted, mark=diamond*] coordinates {
      (100, 0.050095522)
      (200, 0.050106787)
      (300, 0.050117695)
      (400, 0.050128841)
      (500, 0.050139868)
      (600, 0.050150656)
      (700, 0.050161325)
      (800, 0.050171995)
      (900, 0.050182902)
      (1000, 0.050193452)
      };
      \legend{{Checkpointing},
              {Reversibility}
        }
   \end{axis}
\end{tikzpicture}

%% file: figures/vary_steps_check_rev_mem_9.tex
\begin{tikzpicture}
   \begin{axis}[
                xlabel = { Number of Time steps},
                ylabel = {Device Memory Utilization 
                (GiB)}, 
                xtick={100,300,500,700,900},
                ytick={0,0.2, 0.4, 0.8},
                mark options={solid},
                yticklabels={\phantom{0.0}0,0.2, 0.4, 0.8},
                ymin=0,
                legend style={
                    at={(0.25,1.0)},
                    anchor=north,
                    legend columns=1}
                    ]
      \addplot[ultra thick, green, dashed, mark=*] coordinates {
      (100, 0.386477393)
      (200, 0.476345618)
      (300, 0.534966712)
      (400, 0.581873995)
      (500, 0.632687916)
      (600, 0.671785344)
      (700, 0.710885872)
      (800, 0.74634086)
      (900, 0.77372859)
      (1000, 0.812838059)
      };
      \addplot[ultra thick, blue, dotted, mark=diamond*] coordinates {
      (100, 0.195028366)
      (200, 0.195039631)
      (300, 0.195050539)
      (400, 0.195061685)
      (500, 0.195072711)
      (600, 0.1950835)
      (700, 0.195094169)
      (800, 0.195104838)
      (900, 0.195115746)
      (1000, 0.195126296)
      };
      \legend{{Checkpointing},
              {Reversibility}
        }
   \end{axis}
\end{tikzpicture}

%% file: figures/vary_checks_time.tex
\begin{tikzpicture}
   \begin{axis}[
                xlabel = {Checkpoint Interval},
                ylabel = { Execution Time},
                xtick={2,40, 100, 200, 250, 500},
                ymin=0,
                ymax=125,
                mark options={solid},
                legend style={
                    at={(0.25, 0.0)},
                    anchor=south,
                    legend columns=1}
                    ]
      \addplot[ultra thick, red, dotted, mark=*] coordinates {
      (2 , 110.9060) 
(4 , 110.7989) 
(5 , 108.5419) 
(10 , 108.0978) 
(20 , 106.6583) 
(25 , 106.8151) 
(34 , 109.9374) 
(40 , 107.7878) 
(50 , 107.4609) 
(100 , 108.5817) 
(200 , 109.2111) 
(250 , 107.5387) 
(500 , 110.0861)
      };
      \addplot[ultra thick, green, dashed, mark=*] coordinates {
      (2 , 96.1457)
(4 , 94.7103) 
(5 , 95.4159) 
(10 , 95.6859) 
(20 , 93.7549) 
(25 , 94.2336) 
(34 , 94.5100) 
(40 , 93.7590) 
(50 , 95.9777) 
(100 , 95.8358) 
(200 , 97.7988) 
(250 , 94.0504) 
(500 , 94.2148)
      };
      \legend{{Checkpointing},
              {Reversibility}
        }
   \end{axis}
\end{tikzpicture}

%% file: figures/vary_checks_mem.tex
\begin{tikzpicture}
   \begin{axis}[
                xlabel = {Checkpoint Interval},
                ylabel = { Device Memory Utilization},
                xtick={2,40, 100, 200, 250, 500},
                ymin=0,
                mark options={solid},
                legend style={
                    at={(0.25, 1.0)},
                    anchor=north,
                    legend columns=1}
                    ]
      \addplot[ultra thick, red, dotted, mark=*] coordinates {
      (2, 4.093662109)
      (4, 2.189463701)
      (5, 1.810184557)
      (10, 1.086782541)
      (20, 0.812972214)
      (25, 0.793366421)
      (34, 0.820635658)
      (40, 0.851848412)
      (50, 0.929936197)
      (100, 1.437674489)
      (200, 2.570450444)
      (250, 3.148568358)
      (500, 6.043085938)
      };
      \addplot[ultra thick, green, dashed, mark=*] coordinates {
      (2, 4.066318359)
      (4, 2.138682421)
      (5, 1.74768452)
      (10, 0.965688717)
      (20, 0.57830424)
      (25, 0.500104659)
      (34, 0.421905079)
      (40, 0.382805289)
      (50, 0.343705499)
      (100, 0.265505918)
      (200, 0.226406128)
      (250, 0.21858617)
      (500, 0.202946254)
      };
      \legend{{Checkpointing},
              {Reversibility}
        }
   \end{axis}
\end{tikzpicture}